\documentclass{ifacconf}

\usepackage{graphicx}      
\usepackage{natbib}        

\usepackage{epsfig} 
\usepackage{amsmath} 
\usepackage{amssymb}  
\usepackage{bm}
\usepackage{framed}
\usepackage{color}
\usepackage{algorithm}
\usepackage{algorithmicx}
\usepackage{algpseudocode}
\algrenewcommand{\algorithmiccomment}[1]{\hskip3em\% #1}

\newtheorem{assumption}{Assumption}
\newtheorem{definition}{Definition}

\newcommand{\A}{\mathcal{A}}

\newcommand{\f}{\mathsf{f}}

\newcommand{\I}{\mathcal{I}}

\newcommand{\cP}{\mathcal{P}}

\newcommand{\p}{\mathsf{p}}
\newcommand{\R}{\mathbb{R}}
\newcommand{\cR}{\mathcal{R}}

\newcommand{\cS}{\mathcal{S}}

\newcommand{\U}{\mathcal{U}}
\newcommand{\V}{\mathcal{V}}

\newcommand{\Z}{\mathbb{Z}}

\newcommand{\col}{\mathrm{col}}

\newcommand{\supp}[1]{{\rm supp}\left({#1} \right)}

\newcommand{\head}[1]{{\rm head}\left({#1}\right)}
\newcommand{\tail}[1]{{\rm tail}\left({#1}\right)}

\begin{document}
\begin{frontmatter}

\title{Security Index from Input/Output Data: Theory and Computation\thanksref{footnoteinfo}} 

\thanks[footnoteinfo]{This work was supported in part by the Knut and Alice Wallenberg Foundation Wallenberg Scholar, Swedish Research Council (Project 2023-04770), and VINNOVA project ``Control-computing-communication co-design for autonomous industry (3C4AI)'' (No. 2025-01119).}

\author[First]{Takumi Shinohara} 
\author[First]{Karl Henrik Johansson} 
\author[First]{Henrik Sandberg}

\address[First]{Department of Decision and Control Systems, KTH Royal Institute of Technology, and Digital Futures, 100 44 Stockholm, Sweden.\\(E-mail: tashin@kth.se, kallej@kth.se, hsan@kth.se)}

\begin{abstract}                
The concept of a security index quantifies the minimum number of components that must be compromised to carry out a stealth attack.
This metric enables system operators to assess the security risk of each component and implement countermeasures accordingly.
In this paper, we introduce a \textit{data-driven security index} that can be computed solely from input/output data when the system model is unknown.
We show a sufficient condition under which the data-driven security index coincides with the model-based security index, which implies that the exact risk level of each component can be identified solely from data.
We also provide an algorithm for computing the data-driven security index.
\end{abstract}

\begin{keyword}
	Security index, data-driven method, linear systems, security analysis, system security
\end{keyword}

\end{frontmatter}

\section{Introduction}
\label{section:introduction}
%

A key strategy for enhancing cybersecurity and resilience of control systems is the \textit{risk-based approach}, which entails first assessing system risks and then, based on the identified risks, prioritizing and selecting countermeasures accordingly \citep{2015CSMJohansson,2024TACTeixeira}. 
To achieve this, several studies have proposed quantitative metrics to assess the risk level of control systems.
In this paper, we focus on the \textit{security index}.
The original security index, introduced to characterize the vulnerability of sensors in power grids, is defined as the minimum number of sensors that must be compromised while remaining undetected \citep{2010SandbergCPSWEEK,2014TACSandberg}.
While this index was initially proposed for static systems, subsequent studies \citep{2016Sandberg,2019AutomaticaChong} extended this concept to dynamical systems.
In addition, the authors of \citep{2020TACSandberg,2021AutomaticaSandberg} formulated the security index for dynamical systems based on the notion of perfectly undetectable attacks.
For a given component $ i $\footnote{An actuator or a sensor.}, the security index is defined as the minimum number of sensors and actuators that must be compromised to carry out a perfectly undetectable attack involving component $ i $.
Thus, components with a smaller security index are more vulnerable to stealthy attacks than other components.
System operators can assess the risk level of each component based on the security index.

\citep{2020TACSandberg,2021AutomaticaSandberg} introduced security indices for structured systems, addressing cases where system operators possess only incomplete knowledge of the system model.
Yet, as emphasized in the data-driven control literature \citep{2020TACTesi,Data-Driven}, it is often challenging to obtain even the system's structure; instead, often only input/output data (I/O data) are available.
This motivates our central question:
\textit{Given only I/O data, how can we assess the risk level of control systems via a security index?}

In answering this question, our primary contributions can be summarized as follows:
\begin{enumerate}
	\item We develop a data-driven security index which can be computed solely from I/O data, without any model parameters.
	\item We establish a sufficient condition under which the data-driven security index coincides with the model-based security index.
	\item We present an algorithm to compute the data-driven security index.
\end{enumerate}
Several studies (e.g., \citep{2021L-CSSPasquialetti}) have examined data-driven security issues, but most existing research focuses on detecting and identifying attacks.
While our recent work \citep{2026ShinoharaACC} proposed data-driven resilience assessment metrics for sparse sensor attacks, this paper considers a more general problem by incorporating actuator attacks alongside sensor attacks.

The remainder of this paper is organized as follows.
Section \ref{section:problem} presents preliminary descriptions of the system model, data model, attack model, and model-based security index.
In Section \ref{section:data-driven}, we give the definition of the data-driven security index and establish a sufficient condition under which the data-driven and model-based indices coincide.
Additionally, we present an algorithm to compute the index.
This paper concludes in Section \ref{section:conclusion}.

\subsubsection{Notation.}
The symbols $ \R $, $ \R^n$, and $ \mathbb{Z}^+ $ denote the set of real numbers, $ n $-dimensional Euclidean space, and non-negative integers, respectively.
For vectors $ v_1,\ldots,v_k $, define $ \col(v_1,\ldots,v_k) \triangleq [v_1^\top,\ldots,v_k^\top]^\top $.
For a signal $ a:\Z^+ \rightarrow \R^n $, $ a \equiv 0 $ means that $ a(k) = 0$ for all $ k \in \Z^+ $; $ a \not \equiv 0 $ means that $ a(k) \neq 0 $ for at least one $ k \in \Z^+ $.
The notation $ |\I| $ is used to denote the cardinality of a set $ \I $.
For a vector $x$, its support is defined as $ \supp{x} $.
Given a matrix $ A $, we use $ \ker A $ to denote the null space of $ A $.
The identity matrix with size $ n \times n $ is denoted by $ I_n $.
The zero matrix of size $ m \times n $ is denoted by $ 0_{(m,n)} $ and $ 0_m $ is used when the size is $ m \times 1 $ for simplicity.
For $ a, b \in \mathbb{Z}^+ $, we define the integer interval $ [a,b] \triangleq \{c \in \Z^+:a \leq c \leq b \} $. 
Given $ v : \mathbb{Z}^+ \rightarrow \R^n $ and the interval $ [i,j] $, we define a block vector $ v^{[i,j]} $ as $ v^{[i,j]} \triangleq \col(v(i),v(i+1),\ldots,v(j-1),v(j)) $.
For brevity, we sometimes write $ v^{[j]} $ instead of $ v^{[0,j]} $.
Let $ q $ be a positive integer such that $ q \leq j - i + 1 $ and define the Hankel matrix of depth $ q $, associated with $ v^{[i,j]} $, as
\begin{align*}
	\mathcal{H}_q\left(v^{[i,j]}\right) \!\triangleq \!\left[\begin{array}{cccc}
		v(i) & v(i\!+\!1) & \cdots & v(j\!-\!q\!+\!1) \\
		v(i\!+\!1) & v(i\!+\!2) & \cdots & v(j\!-\!q\!+\!2) \\
		\vdots & \vdots & \ddots & \vdots \\
		v(i\!+\!q\!-\!2) & v(i\!+\!q\!-\!1) & \cdots & v(j\!-\!1) \\
		v(i\!+\!q\!-\!1) & v(i\!+\!q) & \cdots & v(j) 
	\end{array}\right]\!.
\end{align*}
Note that the subscript $ q $ refers to the number of block rows of the Hankel matrix.
Then, $ v^{[i,j]} $ is said to be \textit{persistently exciting of order $ q $} if the Hankel matrix $ \mathcal{H}_q(v^{[i,j]}) $ has full row rank \citep{2005SCLWillems,Data-Driven}.
For a Hankel matrix $ H $, we define the head and tail operators, denoted $ \head{H} $ and $ \tail{H} $, which remove the last and first block, respectively, as
\vspace{-1.5mm}
\begin{figure}[h]
	\begin{center}
		\includegraphics[width=0.9\linewidth]{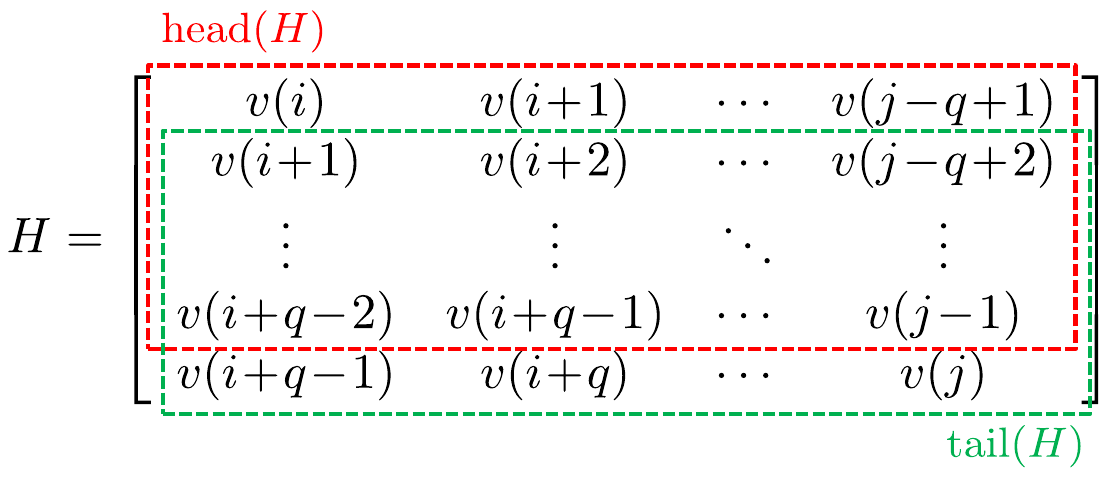}
		\vspace{-3mm}
		\label{fig:head_tail}
	\end{center}
\end{figure}


\section{Problem Formulation}
\label{section:problem}
In this section, we introduce the system model, data representation, and attack model.
We then present the definition and properties of the model-based security index.

\subsection{System Model and Data Representation}
Consider the following linear time-invariant system:
\begin{align}
	\label{eq:sytem_model}
	\left\lbrace \begin{array}{rl}
		x(k+1)\! &= Ax(k) + Bu(k), \\
		y(k)  \!&= Cx(k),
	\end{array} \right.
\end{align}
where $ x(k) \in \R^n $ is the system state, $ u(k)  \in \R^m$ the control input, and $ y(k) \in \R^p $ the sensor output.
We assume that the system is controllable and observable.
Also, $ B $ is assumed to have full column rank.
For notational convenience, define $ \mathcal{M} \triangleq \{1,\ldots,m\} $ and $\mathcal{P} \triangleq \{1,\ldots,p\}$ as the index sets of the inputs and outputs, respectively.

Since we focus on a data-driven setting, we make the following assumption.
\begin{assumption}
	\label{assumption:system}
	The system parameters $ A $, $ B $, and $ C $ are unknown.
	Instead, the I/O data from time $ 0 $ to $ N-1 $ of (\ref{eq:sytem_model}), namely $ u^{[N-1]} $ and $ y^{[N-1]} $, are available, where $ N $ is a sufficiently large integer.
\end{assumption}

By Willems' fundamental lemma \citep{2005SCLWillems}, for a given positive integer $ L $, if $ u^{[N-1]} $ is persistently exciting of order $ n + L $, i.e., the Hankel matrix $ \mathcal{H}_{n+L}(u^{[N-1]}) $ has full row rank, then every $ L $-length I/O trajectory of (\ref{eq:sytem_model}) can always be expressed in terms of $ u^{[N-1]} $ and $ y^{[N-1]} $ as follows:
For some $ k $, $ u^{[k,k+L-1]} $ and $ y^{[k,k+L-1]} $ form an I/O trajectory of (\ref{eq:sytem_model}) if and only if 
\begin{align}
	\label{eq:Willems}
	\left[\begin{array}{c}
		u^{[k,k+L-1]} \\
		y^{[k,k+L-1]}
	\end{array}\right] = 	\left[\begin{array}{c}
		\mathcal{H}_{L}(u^{[N-1]}) \\
		\mathcal{H}_{L}(y^{[N-1]})
	\end{array}\right]g
\end{align}
for some $ g \in \R^{N-L+1} $.

\subsection{Attack Model}
Assume that an adversary can inject malicious signals into both actuators and sensors.
The system under attack is modeled by 
\begin{align}
	\label{eq:sytem_model_attack}
	\left\lbrace \begin{array}{rl}
		x(k+1) \!&= A  x(k) + B u(k)+B_a a(k), \\
		y(k)  \!&= C x(k)+D_a a(k),
	\end{array} \right.
\end{align}
where $ a(k) \triangleq \col(u^a(k),y^a(k)) \in \R^{m+p} $ models the attack designed by the adversary.

Following \citep{2020TACSandberg,2021AutomaticaSandberg}, we make the following assumption on \textit{protected sensors}.
\begin{assumption}
	\label{assumption:protected}
	The last $ \nu \in [0,p] $ elements of the output are protected and cannot be manipulated by the attacker.
\end{assumption}

We denote the set of protected sensors by $ \cS \subseteq \cP $.	
The protection can be achieved by encryption/authentication mechanisms and by improving physical security.
We define the set of unprotected sensors as $ \mathcal{U} \triangleq \cP \backslash \cS $ and the combined set of actuators and unprotected sensors as $ \I \triangleq \{1,\ldots,m+p-\nu\} $.

The first $ m $ entries of $ a $ correspond to attacks against the actuators, while the last $ p $ entries correspond to attacks against the sensors.
We write $y^a$ as $y^a = \col(y^a_\U, y^a_\cS)$, where $ y_\U^a\in\mathbb R^{p-\nu} $ and $ y_\cS^a\in\mathbb R^\nu$ correspond to the unprotected and protected sensors, respectively.
Since the sensors in $\cS$ are protected, we impose $y_\cS^a\equiv 0$.
Consequently, $ B_a $ and $ D_a $ are given by
\begin{align*}
	B_a \triangleq \left[\begin{array}{cc}
		B & 0_{(n,p)}
	\end{array}\right],~
	D_a \triangleq \left[\begin{array}{ccc}
		0_{(p-\nu,m)} & I_{p-\nu} & 0_{(p-\nu,\nu)}  \\ 
		0_{(\nu,m)} & 0_{(\nu,p-\nu)} & 0_{(\nu,\nu)}
	\end{array}\right].
\end{align*}

To define the security index, we first introduce the notion of perfectly undetectable attacks \citep{2020TACSandberg,2021AutomaticaSandberg,2017TCNSSinopoli}.

\begin{definition}
	\label{definition:undetectable_attacks}
	Let $ y(k, x(0), u, a) $ denote the output of the compromised system (\ref{eq:sytem_model_attack}) at time $ k $ generated from the initial condition $ x(0) $, input $ u $, and attack signal $ a \not\equiv 0$.
	The attack signal $ a $ is \textit{perfectly undetectable} if $ y(k, x(0), u, a) \equiv y(k, x(0), u, 0)$.
\end{definition}

Perfectly undetectable attacks are dangerous because the attacker leaves no trace in the measurements.
Consistent with \citep{2020TACSandberg}, one can set $ x(0)=0 $ and $ u \equiv 0 $ without loss of generality, in which case $ y(k, 0,0,a)\equiv 0 $ due to linearity.
This implies that an attack $ a \not\equiv 0 $ is perfectly undetectable if and only if $ y(k,0,0,a) \equiv 0 $.

\subsection{Model-Based Security Index}
Based on perfectly undetectable attacks, the model-based security index $ \delta(i) $ of component $ i \in \I $ is defined as follows \citep{2020TACSandberg,2021AutomaticaSandberg}:
\begin{framed}
\begin{align}
	\label{eq:security_index}
	\hspace{-5mm} \text{(P0):}~~~\delta(i) \triangleq \min_{\{a(k)\}_{k\in\Z^+}} ~\left\| a \right\|_0 
\end{align}
\vspace{-3mm}
\begin{subequations}
	\label{eq:security_index_const}
	\begin{align}
		\mathrm{s.t.} ~~~x(k+1) &= Ax(k) + B_a a(k),  \label{eq:const_1}\\
		0&= Cx(k) + D_a a(k),  \label{eq:const_2}\\
		x(0) &= 0,  \label{eq:const_3}\\
		a_i &\not\equiv 0. \label{eq:const_4}
	\end{align}
\end{subequations}
\end{framed}
Here, $ \|a\|_0 \triangleq |\cup_{k \in \Z^+} \supp{a(k)}|$ is the attack cardinality and $ a_i $ denotes the $ i $th entry of $ a $.
The optimal value equals the minimum number of actuators and unprotected sensors that must be compromised to carry out a perfectly undetectable attack.
The first two constraints (\ref{eq:const_1}) and (\ref{eq:const_2}) ensure that the physical dynamics is consistent with the model under attack, while (\ref{eq:const_2}) and (\ref{eq:const_3}) together ensure that the attack signal is perfectly undetectable.
The last constraint (\ref{eq:const_4}) ensures that component $ i $ is used in the attack.
Note that this optimization problem is posed over all $ k \in \Z^+ $, but in practice, it suffices to check the normal rank condition of the transfer function from $ a $ to $ y $ for a given combination of compromised sensors and actuators.
For the details on the computation of $ \delta(i) $, see \citep[Section III.A]{2020TACSandberg}.

For this index, the following properties were established in \citep{2020TACSandberg,2021AutomaticaSandberg}:
\begin{enumerate}
	\item Components with a smaller $ \delta $ are more vulnerable than those with a larger $ \delta $. The worst case occurs when $ \delta(i)  = 1 $. This value implies that the attacker can attack only component $ i $ and remain perfectly undetectable without compromising other components.
	\item Problem (P0) is not always feasible. Absence of a solution implies that the attacker cannot attack a component $ i $ while remaining perfectly undetectable. In this case, we set $ \delta(i) = + \infty $.
	\item The computation of $ \delta(i) $ is NP-hard. Thus, there are no known polynomial-time algorithms that can be used to solve problem (P0).
\end{enumerate}

The model-based security index is dependent on the numerical entries of the system matrices, but such knowledge is sometimes not available.
Therefore, $ \delta(i) $ is not practical to use in large-scale control systems.
For this issue, a \textit{robust} security index was proposed in \citep{2020TACSandberg} and a \textit{generic} security index was proposed in \citep{2021AutomaticaSandberg}, where neither index requires exact system information and only relies on the system's structural model.
This paper further extends this line of work to more practical scenarios, without assuming the system's structural model; instead, we rely solely on the I/O data.


\section{Data-driven Security Index}
\label{section:data-driven}
In this section, we define the data-driven security index and provide a sufficient condition under which it equals the model-based index.
We then show an algorithm to compute the data-driven security index.

\subsection{Definition}
Given I/O data, protected sensor set $ \cS $, and positive integer $ L $, we define the data-driven security index $ \rho(i) $ of component $ i \in \I $ as 
\begin{framed}
\begin{align}
	\label{eq:data-driven_security_index}
	~\text{(P1):}~~\rho(i) \triangleq \min_{\{g(k)\}_{k \in \Z^+}} \sum_{j  \in \I}\mathbb{I}\left\lbrace \exists k : \left\| L^\f_j g(k)\right\|_2 >0\right\rbrace
\end{align}
\vspace{-3mm}
\begin{subequations}
	\label{eq:data-driven_security_index_const}
	\begin{align}
		\mathrm{s.t.} ~~\left[\begin{array}{c}
			U^\p \\ Y^\p 
		\end{array}\right]g(0)& = 0, \label{eq:const_data_1}\\
		Y^\f_\cS g(k) &= 0,~\forall k \in \Z^+, \label{eq:const_data_2}\\
		Hg(k+1) &= Tg(k),~\forall k \in \Z^+,  \label{eq:const_data_3} \\
		\exists k \in \Z^+ & :\left\| L^\f_i g(k) \right\|_2 > 0. \label{eq:const_data_4} 
	\end{align}
\end{subequations}
\end{framed}
Here, $ g(k) \in \R^{d} $ with $ d\triangleq N-2L+1 $ is the coefficient vector that represents the $ 2L $-length I/O window at time~$ k $ and $ \mathbb{I}\{\cdot\} $ denotes the indicator function, which equals $ 1 $ if its argument is true and $ 0 $ otherwise. 
The matrices $ U^\p,U^\f $ and $ Y^\p, Y^\f $ denote the partitioned I/O Hankel matrices of depth $ 2L $:
\begin{align}
	\left[\begin{array}{c}
		U^\p \\ U^\f
	\end{array}\right] \triangleq \mathcal{H}_{2L}\left(u^{[N-1]}\right),~\left[\begin{array}{c}
		Y^\p \\ Y^\f
	\end{array}\right] \triangleq \mathcal{H}_{2L}\left(y^{[N-1]}\right),
\end{align}
where $ U^\p $ consists of the first $ L $ block rows of the Hankel matrix $ \mathcal{H}_{2L}(u^{[N-1]}) $ and $ U^\f $ consists of the last $ L $ block rows of the Hankel matrix (similarly for $ Y^\p $ and $ Y^\f $).
Refer to $ (U^\p, Y^\p) $ and $ (U^\f, Y^\f) $ as the \textit{past} and \textit{future data}, respectively.
For a given index set $ \Lambda $, $ U^\f_\Lambda $ denotes the submatrix obtained from $ U^\f $ by removing all blocks except those indexed by $ \Lambda $ (similarly for $ Y^\p_\Lambda $ and $ Y^\f_\Lambda $).
Hence, $ Y^\p_\cS $ and $ Y^\f_\cS $ denote, respectively, the submatrices of $ Y^\p $ and $ Y^\f $ indexed by the protected sensors $ \cS $.
For (\ref{eq:const_data_3}), define the shift-consistency matrices $ H, T \in\R^{(m+p)(2L-1)\times d} $ as
\begin{align}
	H \triangleq \left[\begin{array}{c}
		U^\p \\ \head{U^\f} \\ Y^\p \\ \head{Y^\f}
	\end{array}\right],~T \triangleq \left[\begin{array}{c}
		\tail{U^\p} \\ U^\f \\ \tail{Y^\p} \\ Y^\f
	\end{array}\right].
\end{align}
Additionally, regarding (\ref{eq:data-driven_security_index}) and (\ref{eq:const_data_4}), define
\begin{align}
	L^\f_j \triangleq \left\lbrace \begin{array}{ll}
		U^\f_j , &~j \leq m,\\
		Y^\f_{j-m}, &~j>m,
	\end{array} \right.
\end{align}
for given $ j \in \I $.
Recall that $ j \leq m $ indicates that $ j $ is an actuator and $ j > m $ that $ j $ is an unprotected sensor.

Problem (P1) considers a sliding family of $ 2L $-length trajectories that are consistent with the given I/O data and linked by one-step shifts.
The objective function in (\ref{eq:data-driven_security_index}) counts the number of actuators and unprotected sensors whose future blocks are nonzero in at least one window.
The anchor constraint (\ref{eq:const_data_1}) enforces a zero $ L $-length I/O trajectory in the initial window.
Constraint (\ref{eq:const_data_2}) requires all protected sensors to remain zero in every future window.
The shift-consistency constraint (\ref{eq:const_data_3}) requires that each window's past and first $ (L-1) $ future blocks are equal to the previous window's last $ (L-1) $ past and future blocks.
Finally, condition (\ref{eq:const_data_4}) ensures that the target component $ i $ is active in some future window.
The problem is also not always feasible, and in the infeasible case, we set $ \rho(i) = + \infty $.

\subsection{Condition for Equivalence between $ \delta(i) $ and $ \rho(i) $}
The following theorem provides a sufficient condition to ensure the equivalence between $ \delta(i)  $ and $ \rho(i) $.
\begin{thm}
	\label{theorem:equivalent}
	Suppose that Assumptions \ref{assumption:system} and \ref{assumption:protected} hold.
	If $ L \geq n $ and $ u^{[N-1]} $ is persistently exciting of order $ n + 2L $, then $ \delta(i) = \rho(i) $ for every component $ i \in \I $.
\end{thm}
\begin{pf}
We first consider $ \rho(i) \neq + \infty $.
To show $ \rho(i) \geq \delta(i) $, let $ \{g(k)\}_{k \in \Z^+} $ be an optimal solution to the data-driven problem (P1).
From (\ref{eq:const_data_1}), we have $ U^\p g(0) = 0 $ and $ Y^\p g(0) = 0 $, which implies the past $ L $-step input and output blocks are zero.
Since $ L \geq n $ and the system is observable, the finite-time observability matrix of depth $ L $ has full column rank, which implies $ x(0) = 0 $.

Define the $ L $-length trajectory as
\begin{align}
	\label{eq:theorem_03}
	\tilde u^{[k,k+L-1]} \!\triangleq \!U^\f g(k),~\tilde y^{[k,k+L-1]}_\U \!\triangleq \!Y^\f_\U g(k),~\tilde y_{\cS} \!\equiv \!0,
\end{align}
where
\begin{align*}
	\tilde u^{[k,k+L-1]} & \triangleq \col\left(\tilde u(k),\ldots,\tilde u(k+L-1)\right) \in \R^{mL}, \\
	\tilde y^{[k,k+L-1]}_\U & \triangleq \col\left(\tilde y_\U(k),\ldots,\tilde y_\U(k+L-1)\right) \in \R^{|\U| L}, 
\end{align*}
and $ \tilde y_\U(k) \in \R^{|\U|} $ (resp. $ \tilde y_\cS(k) \in \R^{|\cS|} $) is the subvector obtained from $ \tilde y(k) \in \R^p $ by removing all elements except those indexed by $ \U $ (resp. $ \cS $).
Let $ \tilde u(k) = 0 $ and $ \tilde y(k) = 0 $ for all $ k < 0 $.
At $ k = 0 $, by the persistently exciting assumption, Willems' fundamental lemma implies that $ \tilde u^{[0,L-1]} $ and $ \tilde y^{[0,L-1]} $ are an $ L $-length I/O future trajectory of (\ref{eq:sytem_model}).
This means that, for $ k \in [0, L - 1] $, it holds that
\begin{align}
	\label{eq:theorem_03.2}
	x(k+1) = Ax(k) + B \tilde u(k),~\tilde y(k) = Cx(k).
\end{align}
Using definitions of $ B_a $ and $ D_a $ and setting $ \tilde a(k)  \triangleq \col(\tilde u(k), -\tilde y(k))$, these relations are equivalent to 
\begin{align}
	\label{eq:theorem_04}
	x(k+1) = Ax(k) + B_a \tilde a(k),~0 = Cx(k) + D_a \tilde a(k),
\end{align}
for $ k \in [0, L - 1] $.

We then extend these relations to all $ k \in \Z^+ $.
Using (\ref{eq:const_data_3}) and $ U^\p g(0) = 0 $, by the recursive property, we have
\begin{align*}
	U^\p g(k) = \left\lbrace \begin{array}{ll}
		\col(0_{m(L-k)}, \tilde u^{[0,k-1]}),&~k < L, \\
		\tilde u^{[k-L, k-1]}, &~k\geq L.
	\end{array} \right.
\end{align*}
A similar relation can be derived for $ Y^\p g(k) $.
Then, from (\ref{eq:theorem_03}), we have 
\begin{align*}
	\left[\begin{array}{c}
		U^\p \\ U^\f
	\end{array}\right]g(k) &=\left[\!\begin{array}{c}
		\tilde u^{[k-L,k-1]} \\\tilde u^{[k,k+L-1]}
	\end{array}\right] = \tilde u^{[k-L,k+L-1]},\\
	\left[\begin{array}{c}
	Y^\p \\ Y^\f
\end{array}\right]g(k) &=\left[\!\begin{array}{c}
	\tilde y^{[k-L,k-1]} \\\tilde y^{[k,k+L-1]}
\end{array}\right] = \tilde y^{[k-L, k+L-1]},
\end{align*}
for all $ k \in \Z^+ $.
Since the input is persistently exciting of order $ n+2L $, by Willems' fundamental lemma, $ 2L $-length trajectories $ \tilde u^{[k-L,k+L-1]} $ and $ \tilde y^{[k-L, k+L-1]} $ are the I/O trajectories of the system (\ref{eq:sytem_model}).
Therefore, for all $ k\in \Z^+ $, (\ref{eq:theorem_03.2}) and (\ref{eq:theorem_04}) hold and the sequence $ \tilde a \triangleq \col(\tilde u, -\tilde y) $ is a perfectly undetectable attack, which implies $ \delta(i) \leq \| \tilde a\|_0 $.
Moreover, by (\ref{eq:const_data_3}), adjacent windows match on their $ (2L-1) $-block overlap (via head/tail operator), so these windows can be stitched into a single infinite trajectory $ (\tilde u, \tilde y) $.
From the objective (\ref{eq:data-driven_security_index}), it follows that
\begin{align*}
	\left\| \tilde a \right\|_0 &\!=\! \left\| \tilde u \right\|_0 \!+\! \left\| \tilde y \right\|_0 \!=\!\left| \bigcup_{k \in \Z^+} \supp{\tilde u(k)}\right|\!+\!\left| \bigcup_{k \in \Z^+} \supp{\tilde y(k)}\right| \\
	& \!=\!\sum_{j  \in \mathcal{I}} \mathbb{I}\left\lbrace \exists k : \left\| L^\f_j g(k)\right\|_2 >0  \right\rbrace = \rho(i),
\end{align*}
which implies $ \delta(i) \leq \rho(i) $.

Next, we show $ \delta(i)\geq \rho(i) $.
Let $ a \triangleq \col(u^a, y^a) $ be an optimal solution of the model-based security index problem (P0) for component $ i $.
Since protected sensors $ \cS $ are not attackable, $ y^a_\cS \equiv 0 $.
By constraints (\ref{eq:const_1})--(\ref{eq:const_3}), there exists a state sequence $ \{x(k)\}_{k \in \Z^+} $ such that, with input $ \{a(k)\}_{k \in \Z^+} $ and initial state $ x(0) = 0 $,
\begin{align}
	x(k+1) = Ax(k) + B_a a(k),~0=Cx(k) + D_a a(k),
\end{align}
for all $ k \in \Z^+ $.
By the structure of $ B_a $ and $ D_a $, and since $y_\cS^a\equiv0$, this can be written as
\begin{align}
	\label{eq:theorem_05}
	x(k+1)=Ax(k) + Bu^a(k),~-y^a(k) = Cx(k),
\end{align}
which indicates that $ u^a $ and $ -y^a $ act as an ordinary input and output of the original system (\ref{eq:sytem_model}), respectively.
For notational convenience, extend $u^a(k)$, $y^a(k)$, and $x(k)$ by zero for all $k<0$.
This zero extension is consistent with (\ref{eq:theorem_05}) because $x(0)=0$ and the attack is zero before $k_0$.
Then, for $ k \in \Z^+ $, consider the $ 2L $-length trajectory:
\begin{align*}
	\tilde u^a(k) &\triangleq(u^a)^{[k_0+k-L, k_0+k+L-1]}, \\
	\tilde y^a(k) &\triangleq (y^a)^{[k_0+k-L, k_0+k+L-1]},
\end{align*}
where $ k_0 \in \Z^+ $ is the first time with $ a(k_0) \neq 0 $ (i.e., the first attack occurs).
Then, from (\ref{eq:theorem_05}), $ \tilde u^a(k) $ and $ - \tilde y^a(k) $ are $ 2L $-length I/O trajectory of the system (\ref{eq:sytem_model}), respectively.
Using the persistently exciting assumption and Willems’ fundamental lemma, this trajectory can be represented by the data.
In particular, for all $ k \in \Z^+ $, there exist $ g(k) \in \R^{d} $ such that
\begin{align}
	\label{eq:theorem_06}
	\left[\begin{array}{c}
		\tilde u^a(k) \\ - \tilde y^a(k)
	\end{array}\right] = 
	\left[\begin{array}{c}
		U^\p \\U^\f \\  Y^\p \\  Y^\f
	\end{array}\right]g(k).
\end{align}
From the definition of $ k_0 $, for $ k = 0 $, we have $ U^\p g(0) = 0 $ and $ Y^\p g(0) = 0$, which implies that (\ref{eq:const_data_1}) holds.
Also, given that $ y^a_{\cS} \equiv 0 $, the corresponding future blocks satisfy $ Y^\f_\cS g(k) = 0 $ for all $ k \in \Z^+ $, namely (\ref{eq:const_data_2}) holds.
From (\ref{eq:theorem_06}), $ \{g(k)\} $ represents consecutive $ 2L $-length sliding windows of the same I/O trajectory.
Therefore, the head-tail relations across adjacent windows are identical by construction, which is precisely (\ref{eq:const_data_3}).
Additionally, (\ref{eq:const_4}) implies $ u^a_i \not\equiv 0 $ (if component $ i $ is an actuator) or $ y^a_{i-m} \not\equiv 0 $ (if component $ i $ is a sensor), which yields (\ref{eq:const_data_4}).
Therefore, $ \{g(k)\}_{k \in \Z^+} $ is feasible for the data-driven problem (P1).
From (\ref{eq:data-driven_security_index}),
\begin{align*}
	\rho(i) & \leq \sum_{j  \in \mathcal{I}} \mathbb{I}\left\lbrace \exists k : \left\| L^\f_j g(k)\right\|_2 >0  \right\rbrace\nonumber \\
	& = \sum_{t  \in \mathcal{M}} \mathbb{I}\left\lbrace \exists k \!:\! \left\| \tilde u^a_t(k)\right\|_2 \!>\!0 \right\rbrace \!+\! \sum_{\ell \in \mathcal{U}} \mathbb{I}\left\lbrace \exists k \!:\! \left\| \tilde y^a_\ell (k) \right\|_2 \!>\!0 \right\rbrace\\
	& = \left\| u^a \right\|_0 + \left\| y^a \right\|_0 = \delta(i),
\end{align*}
where $ \tilde u^a_t(k) $ and $ \tilde y^a_\ell(k) $ are the subvectors obtained from $ \tilde u^a(k) $ and $ \tilde y^a(k) $ by removing all entries except those related to component $ t $ and $ \ell $, respectively.
Together with the previous result, this gives $ \delta(i) = \rho(i) $ if $ \rho(i) \neq +\infty $.

Next, consider $ \rho(i) = +\infty $.
This implies that (P1) is infeasible and there is no optimal solution $ \{g(k)\}_{k \in \Z^+} $.
Assume, for contradiction, that $ \delta(i) < +\infty $.
Then, there exists a perfectly undetectable attack $ a \triangleq \col(u^a,y^a) $ using component $ i $.
The input and output attack sequences $ u^a $ and $ y^a $ follow (\ref{eq:theorem_05}).
Let $ k_0 $ be its first nonzero time, and consider the $ 2L $-length intervals $ \tilde u^a(k) $ and $ \tilde y^a(k) $ as defined above.
Under the persistently exciting assumption, Willems' fundamental lemma implies that these $ 2L $-length trajectories are represented by the I/O data, i.e., (\ref{eq:theorem_06}) holds for some $ g(k) \in \R^{d} $.
Then, from the aforementioned analyses, we know that there exists $ \{g(k)\}_{k \in \Z^+} $ satisfying (\ref{eq:const_data_1})--(\ref{eq:const_data_4}), but this contradicts the premise that $ \rho(i) = + \infty $.
Therefore, $ \delta(i) = +\infty $, and thus $ \delta(i) = \rho(i) $. \qed
\end{pf}

This theorem implies that, under the conditions that $ L \geq n $ and $ u^{[N-1]} $ is persistently exciting of order $ n+2L $, the data-driven security index coincides with the model-based security index.
Consequently, one can precisely assess the risk level of each component by solving (P1) using only the persistently exciting data.
The optimization variable $ \{g(k)\}_{k \in \Z^+} $ in Problem (P1) is an infinite sequence, but one can compute $ \rho(i) $ with a finite number of steps, as described in the next subsection.

\subsection{Computation of $ \rho(i) $}

For a set $ \V \subseteq \R^{d} $, we define the pre- and post-operators as
\begin{align}
	\label{eq:set_pre}
	\mathrm{Pre}(\V)  &\triangleq \left\lbrace g \in\R^d: \exists v \in \V~\text{s.t.}~Hv = Tg\right\rbrace,\\
	\label{eq:set_post}
	\mathrm{Post}(\V) &\triangleq \left\lbrace v\in\R^d: \exists g \in \V~\text{s.t.}~Hv = Tg\right\rbrace.
\end{align}
For a set $ \Gamma \!\subseteq\! \I $, define the sequence $ \{\V_t\}_{t \in \Z^+} \!\!\subseteq \!\R^d$ as\footnote{The sequence $ \{\V_t\}_{t \in \Z^+} $ (also $ \{\cR_t\}_{t \in \Z^+} $) depends on $ \Gamma \subseteq \I $, but we omit indicating $ \Gamma $ in this subsection for simplicity.}
\begin{align}
	\label{eq:V_set}
	\V_{t+1} \!\triangleq \!\V_0 \cap \mathrm{Pre}(\V_t) ,~\V_0 \!\triangleq\! \ker \left[\begin{array}{c}
		Y^\f_\cS \\ U^\f_{\bar \Gamma_u} \\ Y^\f_{\bar \Gamma_y}
	\end{array}\right]\!\!,~\V_\infty \!\triangleq\! \lim_{t \rightarrow \infty}\! \V_t,
\end{align}
where 
\begin{align}
	\label{eq:Gamma_u}
	\Gamma_u &\triangleq \mathcal{M} \cap \Gamma, &\bar \Gamma_u &\triangleq \mathcal{M}\backslash \Gamma_u, \\
	\label{eq:Gamma_y}
	\Gamma_y &\triangleq \{\ell \in \U: m+\ell \in \Gamma \},&\bar \Gamma_y &\triangleq \mathcal{U}\backslash \Gamma_y.
\end{align}
Hence, $ \bar \Gamma_u $ and $ \bar \Gamma_y $ denote the index sets of actuators and unprotected sensors not included in $ \Gamma $, respectively.
By construction, one can obtain that $ \V_\infty \subseteq \cdots \subseteq \V_{1} \subseteq \V_0 $.
Also, it holds that $ \V_\infty \subseteq \mathrm{Pre}(\V_\infty) $.
Additionally, define the sequence $ \{\cR_t\}_{t \in \Z^+} \subseteq \R^d $ as
\begin{align}
	\label{eq:R_set}
	\cR_{t+1} &\triangleq \V_\infty \cap \left(\cR_t + \mathrm{Post}(\cR_t)\right),~\cR_0 \!\triangleq\V_\infty \cap \ker \left[\begin{array}{c}
		U^\p \\ Y^\p
	\end{array}\right],\nonumber \\
	\cR_\infty &\triangleq \bigcup_{t=0}^{\infty} \cR_t,
\end{align}
where $+$ denotes the sum of subspaces.
By construction, we have $ \cR_0 \subseteq \cR_1 \subseteq \cdots \subseteq \cR_\infty $.

\begin{rem}
	\label{remark:V_and_R}
For given $ \Gamma \subseteq \I $, the sequence $ \{\V_t\}_{t \in \Z^+} $ consists of subspaces and characterizes the set of parameters that can keep satisfying the zero constraints and the one-step shift consistency indefinitely.
Since $\{\V_t\}$ is a decreasing sequence of subspaces in $\R^d$, it converges in at most $d$ steps.
The sequence $\{\cR_t\}_{t\in\Z_+}$ is an increasing sequence of subspaces in $\R^d$ contained in $\V_\infty$.
It is the linear span of finite reachable terminal points from the initial anchor (\ref{eq:const_data_1}) while remaining in $\V_\infty$.
The integer sequence $\{\dim \cR_t\}$ is monotone and bounded by $d$.
Hence $\{\cR_t\}$ converges in at most $d$ steps.
Therefore, both $\V_\infty$ and $\cR_\infty$ can be computed by at most $d$ iterations.
\end{rem}

To present an algorithm for computing $ \rho(i) $, we derive the following lemma.
\begin{lem}
	\label{lemma:equivalence}
	Suppose that Assumptions \ref{assumption:system} and \ref{assumption:protected} hold.
	For component $ i \in \I $ and a subset $ \Gamma \subseteq \I $ which contains $ i $, the following statements are equivalent:
	\begin{enumerate}
		\item [(i)] There exists $ \{g(k)\}_{k \in \Z^+} $ satisfying (\ref{eq:const_data_1})--(\ref{eq:const_data_4}) and $ g(k) \in \V_0 $ for all $k \in \Z^+$. 
		\item [(ii)] $ \cR_\infty \not\subseteq \ker L^\f_i $.
	\end{enumerate}
	Additionally, for any $ \{g(k)\}_{k \in \Z^+} $ satisfying (i), it holds that $ g(k) \in \V_\infty,~\forall k \in \Z^+ $.
\end{lem}
\begin{pf}
	For (i) $ \Rightarrow $ (ii), let $ \{g(k)\}_{k \in \Z^+} $ satisfy (i).
	From (\ref{eq:const_data_3}), we have $ Hg(k+1) = Tg(k),~\forall k \in \Z^+ $.
	Based on the definition of $ \mathrm{Pre}(\cdot)$ and the assumption that $ g(k) \in \V_0 $ for all $k \in \Z^+$, $ g(k) \in \mathrm{Pre}(\V_0),~\forall k \in \Z^+ $, and thus,
	\begin{align*}
		g(k) \in \V_0 \cap \mathrm{Pre}(\V_0) = \V_1,~\forall k \in \Z^+.
	\end{align*}
	The equality follows by the first equality in (\ref{eq:V_set}).
	Recursively, we have $ g(k)\in\mathrm{Pre}(\V_1),~\forall k \in \Z^+ $, which yields
	\begin{align*}
		g(k) \in \V_0 \cap \mathrm{Pre}(\V_1) = \V_2,~\forall k \in \Z^+.
	\end{align*}
	Therefore, by induction, we have $ g(k) \in \V_\infty $ for all $ k \in \Z^+ $. 
	Moreover, since $g(0)\in \V_\infty$ and (\ref{eq:const_data_1}) holds, we have
	\begin{align*}
	g(0)\in \V_\infty\cap\ker
	\begin{bmatrix}
		U^\p\\
		Y^\p
	\end{bmatrix}
	=\cR_0.	
	\end{align*}
	If $g(k)\in \cR_k$, then $Hg(k+1)=Tg(k)$ implies $g(k+1)\in\mathrm{Post}(\cR_k)$.
	Since $g(k+1)\in \V_\infty$,
	we have
	\begin{align*}
		g(k+1)\in \V_\infty\cap\mathrm{Post}(\cR_k) \subseteq \cR_{k+1}.
	\end{align*}
	Thus $g(k)\in \cR_k\subseteq \cR_\infty$ for all $k$.
	By (\ref{eq:const_data_4}), there exists $k^*$ such that $L_i^\f g(k^*)\neq0$.
	Therefore, $\cR_\infty\not\subseteq\ker L_i^\f$, which implies that (ii) holds.
	
	
	Next, for (ii) $ \Rightarrow $ (i), suppose that (ii) holds.
	By induction on $t$, $\cR_t$ is the linear span of all terminal points of finite sequences of length at most $t$ in $\V_\infty$ starting from $\cR_0$.
	This follows from $\V_\infty\subseteq\mathrm{Pre}(\V_\infty)$ and $0\in \cR_0$.
	Since $\cR_\infty \not\subseteq \ker L^\f_i$, there exists $\gamma\in \cR_\infty$ such that $L_i^\f\gamma\neq0$.
	By the above span characterization, $\gamma$ is a finite linear combination of such terminal points. 
	Since $\ker L_i^\f$ is a subspace, at least one of these terminal points, denoted by $\gamma^*$, must satisfy $L_i^\f\gamma^*\neq0$.
	Hence, there exists a finite sequence
	$\{\gamma(k)\}_{k=0}^s\subset \V_\infty$ such that
	\begin{align*}
	\gamma(0)\in \cR_0,~~H\gamma(k+1)=T\gamma(k),~~k=0,\ldots,s-1,	
	\end{align*}
	and
	\begin{align*}
	\gamma(s)=\gamma^*,~~L_i^\f\gamma(s)\neq0.
	\end{align*}
	Since $\V_\infty\subseteq\mathrm{Pre}(\V_\infty)$, this finite sequence can be extended indefinitely while remaining in $\V_\infty$.
	Namely, for each $k\ge s$, choose $\gamma(k+1)\in \V_\infty$ such that $H\gamma(k+1)=T\gamma(k)$.
	The resulting infinite sequence satisfies (\ref{eq:const_data_3}) by construction. 
	Since $\gamma(0)\in \cR_0$, it satisfies (\ref{eq:const_data_1}).
	Since $\gamma(k)\in \V_\infty\subseteq \V_0$ for all $k$, it satisfies $g(k)\in \V_0$ and hence (\ref{eq:const_data_2}).
	Finally, $L_i^\f\gamma(s)\neq0$, so (\ref{eq:const_data_4}) holds. Therefore, (i) holds.
	\qed

\end{pf}

This lemma shows that, for a fixed $\Gamma\subseteq\I$ containing $i$, condition (ii) holds if and only if there exists a sequence satisfying the constraints of (P1) whose active components are contained in $\Gamma$.
By this lemma, the following proposition on computing the data-driven security index is given.
\begin{prop}
	\label{proposition:rho}
	Suppose that Assumptions \ref{assumption:system} and \ref{assumption:protected} hold.
	The data-driven security index is given by
	\begin{align}
		\label{eq:proposition_rho}
		\rho(i)=\min \left\lbrace |\Gamma|:\Gamma\subseteq \I,~i \in \Gamma,~\cR_\infty \not\subseteq \ker L^\f_i \right\rbrace.
	\end{align}
\end{prop}
\begin{pf}
	Let $ \Gamma^* $ be a minimizer of the right-hand side of (\ref{eq:proposition_rho}).
	By Lemma~\ref{lemma:equivalence}, condition (i) holds for $ \Gamma^* $, and thus, there exists $ \{g^*(k)\}_{k \in \Z^+} $ such that (\ref{eq:const_data_1})--(\ref{eq:const_data_4}) hold and $ g^*(k) \in \V_0 $ for all $k \in \Z^+$.  
	Then, from the definition of $ \V_0 $, for all $ g^*(k) $, it follows that
	\begin{align*}
		\left\lbrace \begin{array}{ll}
			U_j^\f g^*(k) = 0,&~\forall j \in \bar \Gamma^*_u, \\
			Y_\ell^\f g^*(k) = 0,&~\forall \ell \in \bar \Gamma^*_y.
		\end{array}\right.
	\end{align*}
	Consequently, by the objective in (\ref{eq:data-driven_security_index}), we have 
	\begin{align*}
		\rho(i) &\leq \sum_{j  \in \mathcal{I}} \mathbb{I}\left\lbrace \exists k : \left\| L_j^\f g^*(k)\right\|_2 >0  \right\rbrace \leq |\Gamma^*|.
	\end{align*}
	Conversely, let $ \{g(k)\}_{k\in \Z^+} $ be an optimal solution of (P1).
	Define 
	\begin{align}
		\mathcal{A}\triangleq \left\lbrace j \in \I:\exists k \in \Z^+~\mathrm{s.t.}~L_j^\f g(k) \neq 0 \right\rbrace.
	\end{align}
	From the objective in (\ref{eq:data-driven_security_index}), it follows that $ \rho(i) = |\mathcal{A}| $ and 
	\begin{align*}
		g(k) \in \ker \left[\begin{array}{c}
			Y^\f_\cS \\ U^\f_{\bar \A_u} \\ Y^\f_{\bar \A_y}
		\end{array}\right],~\forall k \in \Z^+,
	\end{align*}
	where $ \bar \A_u $ and $ \bar \A_y $ are defined based on $ \A $ similarly to $ \bar \Gamma_u $ and $ \bar \Gamma_y $, respectively (cf. (\ref{eq:Gamma_u}) and (\ref{eq:Gamma_y})).
	Hence, for $ \mathcal{A} $, (i) of Lemma~\ref{lemma:equivalence} holds, equivalently, (ii) holds. 
	This implies that $ \mathcal{A} $ is a feasible set of the right-hand side of (\ref{eq:proposition_rho}).
	Therefore, $ |\Gamma^*| \leq |\mathcal{A}| = \rho(i) $, which concludes the proof.	\qed
\end{pf}

This proposition implies that the data-driven security index $ \rho(i) $ can be computed based on $ \cR_{\infty} $, which can be derived using the sequences $ \{\V_k\} $ and $ \{\cR_k\} $.
Recalling Remark~\ref{remark:V_and_R}, for each fixed $\Gamma$, the sequences $ \{\V_k\} $ and $ \{\cR_k\} $ converge in at most $ d $ iterations.
This leads to Algorithm~\ref{algorithm:rho}, which computes the data-driven security index using data.

\begin{algorithm}[t]
	\caption{Computing data-driven security index $ \rho(i) $}
	\label{algorithm:rho}
	\begin{algorithmic}[1]
		\Statex \hspace{-6mm} \textbf{Input:} $ U^\p, Y^\p, U^\f, Y^\f, \I, \cS, d, L^\f_i $, and $ i $
		\Statex \hspace{-6mm} \textbf{Output:} $ \rho(i) $
		\For{$  q = 1,\ldots,|\I| $}
		\For{all $ \Gamma \subseteq \I $ such that $ |\Gamma| = q $ and $ i \in \Gamma $}
		\State Compute $ \V_0 $ based on (\ref{eq:V_set}).
		\For{$ t = 0,\ldots,d-1 $}
		\State $ \V_{t+1} \leftarrow \V_0 \cap \mathrm{Pre}(\V_t) $.
		\EndFor
		\State Set $ \V_\infty \leftarrow \V_d $.
		\State Compute $ \cR_0 $ based on (\ref{eq:R_set}). 
		\For{$ t = 0,\ldots,d-1 $}
		\State $ \cR_{t+1} \leftarrow \V_\infty \cap \left(\cR_t + \mathrm{Post}(\cR_t)\right) $.
		\EndFor
		\State Set $ \cR_\infty \leftarrow \cR_d $.
		\If{$ \cR_\infty \not\subseteq \ker L^\f_i $}
		\State \textbf{return} $ \rho(i)  = q $ 
		\EndIf
		\EndFor
		\EndFor
		\State \textbf{return} $ \rho(i)  = +\infty $ 
	\end{algorithmic}
\end{algorithm}

\section{Conclusion}
\label{section:conclusion}
We introduced a data-driven security index $ \rho(i) $, which can be computed solely from I/O data, without any model parameters.
We proved a condition under which $ \rho(i) $ coincides with the model-based index $ \delta(i) $.
Under the conditions of Theorem~\ref{theorem:equivalent}, one can obtain the exact security index of each component from data via Algorithm~\ref{algorithm:rho}.
Future work will address noisy I/O data, including robustness and finite-sample effects.

\section*{Usage of Generative AI}
The authors used ChatGPT throughout the manuscript to improve its grammar and clarity.
After using this tool, the authors reviewed and edited the content and take full responsibility for the publication’s content.

{
\bibliography{ifacconf.bib}}             
\end{document}